# Ultrathin and highly passivating silica shells for luminescent and water-soluble CdSe/CdS nanorods

Xiao Tang, Elvira Kröger, Andreas Nielsen, Christian Strelow, Alf Mews, and Tobias Kipp *

*Institute of Physical Chemistry, University of Hamburg, Grindelallee 117, 20146 Hamburg, Germany*

## Abstract

Microemulsion (water-in-oil) methods enable the encapsulation of individual nanoparticles into $SiO_2$ spheres. The major drawbacks of this method, when applied for silica encapsulation of anisotropic nanorods (NRs), are a spatially unequal silica growth and long reaction times (24 h at least). In this work, various tetra-alkoxysilanes (tetramethyl orthosilicate (TMOS), tetraethyl orthosilicate (TEOS) and tetrapropyl orthosilicate (TPOS)) with different alkyl-chain lengths were used as the silica precursors in attempt to tune the silanization behavior of CdSe/CdS NRs in a microemulsion system. We find an enhanced spatial homogeneity of the silica growth with decreasing alkyl-chain length of the tetra-alkoxysilanes. In particular, by using TMOS as the precursor, NRs can be fully encapsulated in a continuous thin ($\leq 5$ nm) silica shell within only 1 h reaction time. Surprisingly, the thin silica shell showed a superior shielding ability to acidic environment even compared to the 30 nm thick shell prepared by using TEOS. Our investigations suggest that the lower steric hindrance of TMOS compared to TEOS or TPOS strongly promotes the homogeneous growth of the silica shells while its increased hydrolysis rate leads decreases the porosity of these shells.

## Introduction

In the past three decades, colloidal II-VI semiconductor nanoparticles (NPs) have attracted significant interest owing to their unique characteristics such as high quantum efficiency, size-tunable fluorescence and solution processability.[1] One of the most promising applications of the NPs is their use in biological sensing.[2] With this objective, it is advantageous to encapsulate the NPs with silica shells since silica does not only ensure the NPs' aqueous dispersibility and biocompatibility, but can also prevent degradation of NPs in oxygen-rich or acidic environments.[3–5] Generally, the preparative methods of growing silica shells can be divided into two strategies, which are the conventionally used Stöber method[6–10] and the lately developed microemulsion (water-in-oil) method.[4,11–17] As an alternative to the Stöber method, the microemulsion method shows its advantages in being less harsh and in terms of forming individual $SiO_2$ spheres encapsulating single NPs.[9,13] The key for creating such a microemulsion system is to use nonionic tensides (usually poly (5) oxyethylene- 4- nonylphenyl- ether (Igepal CO-520)) that form reversed micelles, which can provide a water-based environment between the hydrophobic surface of NPs and the continuous organic solvent phase (usually cyclohexane). Within this water layer, silica precursor (usually tetraethyl orthosilicate (TEOS)) is hydrolyzed, with ammonia acting as a catalyst (eq (1)). The hydrolyzed TEOS species then nucleates before a $\equiv$Si–O–Si$\equiv$ network builds up with other hydrolyzed or unhydrolyzed TEOS species via hydrolytic (eq. (2)) or alcoholic (eq. (3)) polymerization, letting the silica shell grow. Because the hydrolysis of TEOS is efficiently restricted to the water layer closely surrounding the NPs, silica shells with uniform thickness can be routinely formed around individual NPs.[12]

$$Si(OC_2H_5)_4 + xH_2O \rightarrow Si(OC_2H_5)_{4-x}(OH)_x + xC_2H_5OH \quad (1)$$

$$\equiv Si-OH + HO-Si\equiv \rightarrow \equiv Si-O-Si\equiv + H_2O \quad (2)$$

$$\equiv Si-OH + H_5C_2O-Si\equiv \rightarrow \equiv Si-O-Si\equiv + C_2H_5OH \quad (3)$$

Compared to spherical NPs, it is challenging to apply this method for anisotropic nanorods (NRs). It is believed that the high curvature on the two tips leads to a less compact distribution of the attached organic ligands (including the original phosphonic acids and the later-introduced tensides), and therefore to a lower steric hindrance compared to that on the lateral facets.[18–20] Also, different facets induce different stabilities of the micellar surrounding and exhibit different reactivities.[4,13] As a result, during the silanization process, TEOS molecules can typically more easily attach on the two tips of the NRs.[4] With the reaction proceeding, the nucleated silica species first grow into two distinct spheres, to form a dumbbell-like structure; then with the size increasing, the two spheres eventually unify to form a large (thickness ≥ 20 nm) silica ellipsoid. Here, the minimum thickness (likely on the waist) of the finally formed continuous silica shell on a NR is largely limited by the required radius, with which the distinct two spheres can unify.[4] In order to coat NRs with uniform thin silica shells, it is beneficial to enhance the spatial homogeneity of the whole silanization process. Very recently, Hutter et al. and Anderson et al. demonstrated the successful formation of ultrathin (≤ 5 nm) silica shells on NRs by optimizing the relative volume of water and ammonia in the microemulsion system, respectively.[11,13] However, in their research, very long reaction times (from 24 hours to 1 week) were required for the entire $SiO_2$ encapsulation.

Here, we report on the fundamental modification of the silanization dynamics on the NRs' surface by utilizing other silica precursors as alternatives to the conventionally used TEOS. It has been demonstrated that the alkyl-chain length governs the reactivity of tetra-alkoxysilanes.[21] Thus, we reasonably assume that the alkyl-chain length can also efficiently influence the substituting ability of tetra-alkoxysilanes. Keeping this in mind, in the present work, we used tetramethyl orthosilicate (TMOS), TEOS, and tetrapropyl orthosilicate (TPOS) as the precursor to silanize 50 nm long CdSe/CdS NRs. The alkyl-chain length increases from TMOS over TEOS to TPOS, as methoxy, ethoxy, and propoxy groups are connected to the center silica atom in the corresponding molecules, respectively. We find that the use of TMOS can lead to uniform thin (≤ 5 nm) silica shells around the NRs within reaction times of only 1 hour, which is much shorter than in the aforementioned previous studies.[11,13] The TMOS-derived thin shells show superior photoluminescence properties in comparison to the conventional TEOS-derived thick (20 nm) silica shells. In particular, the TMOS-prepared structures show a more robust shielding ability to acidic environment, which can be attributed to the lower porosity of the thin silica shells.

## Experimental Section

**Synthesis of CdSe/CdS NRs:** CdSe/CdS NRs were fabricated according to the method described by Carbone and coworkers.[22] First, CdSe dots were prepared prior to the epitaxial growth of CdS rods. The CdSe precursor solution was prepared by mixing CdO (0.051 g), octa-decyl-phosphonic acid (ODPA) (0.29 g) and trioctylphosphine oxide (TOPO) (3.0 g) in a 50 mL three-neck flask. The mixture was kept under vacuum at 110 °C for 1 hour to remove crystallized water. Afterwards, the mixture was heated to 300 °C under nitrogen and reflux conditions for 10 min until the liquid mixture became colorless and transparent. Then the mixture was naturally cooled to 250 ℃. At this point 1.6 mL Se-tributylphosphine (TBP) solution (0.5 M) was rapidly injected under vigorous stirring. After 3 minutes, the heating mantle was removed and the mixture was cooled to room temperature by using a fan. 10 mL methanol was added to totally quench the reaction. The dots were separated by centrifugation at 11000 rpm. After that, the dots were dispersed in 5 mL trioctylphosphine (TOP) and stored in a glovebox under inert gas atmosphere. To prepare core/shell CdSe/CdS NRs, firstly, CdO (0.086 g), ODPA (0.29 g), hexylphosphonic acid (HPA) (0.08 g) and TOPO (3 g) were mixed in a 50 mL three-neck flask. The mixture was kept under vacuum at 150 ℃ for 2 hours to remove crystallized water. After that, refluxed by cooling water, the mixture was heated up to 370 ℃ under nitrogen. When the temperature was stabilized, TOP (0.9 mL) was rapidly injected. The flask was then cooled to 350 ℃, at which a mixture of already-prepared 0.8 mL CdSe-dot solution, 0.6 mL TOP and 0.8 mL sulphur-TOP solution (2 M) was rapidly injected to trigger the rod-growth reaction. The reaction was kept for 12 minutes accurately, after that, the heating mantle was removed and the mixture was cooled to room temperature by using a fan. 10 mL methanol was added to totally quench the reaction. The NRs were separated by centrifugation at 11000 rpm and finally dispersed in 10 mL cyclohexane. Transmission electron microscopy images indicate widths and lengths of the NRs used in this work of about 5 nm and 50 nm, respectively.

**Silanization of CdSe/CdS NRs:** The silanization of the NRs was performed in glass vials at room temperature. In each vial, 380 µL cyclohexane dispersion of NRs was added into a mixture of 2 mL Igepal CO-520 and 30 mL cyclohexane. After 2 hours of stirring, 100 µL ammonia aqueous solution (25%) was added. At this point, various tetra-alkoxysilanes as the silica precursor were separately introduced to each vial to start the silanization process. In one vial, a rather large volume of TEOS (400 µL) was introduced to grow a thick $SiO_2$ shell as a reference sample. In attempt to grow thin $SiO_2$ shells, much lower volumes of TMOS, TEOS and TPOS (60 µL, 82 µL and 104 µL, respectively, corresponding to equal molar amounts) were separately added to three other vials. After the addition of the tetra-alkoxysilanes, all these mixtures were stirred for different periods of time (1 h to 24 h). To quench the reactions, 20 mL ethanol was added into each mixture. The solids were then separated from the liquid by 11000 rpm centrifugation and re-dispersed in 30 mL ethanol with the help of 5 min of ultrasonication.

**ATR-FTIR spectroscopy:** Infrared-absorption characterizations were carried out on a Varian 660 FTIR spectrophotometer with a resolution of 4 $cm^{-1}$. Spectra of Igepal CO-520 were directly recorded on the purchased chemical without further treatment. To record spectra of pure NRs, Igepal-functionalized NRs, as well as TMOS-, TEOS- and TPOS-treated NRs, the according reaction mixtures were all washed with ethanol and dispersed in ethanol at the same concentration of NRs. Then 10 µL of each solution were dropped onto the sample holder. After that, samples were exposed to nitrogen flow for 15 min to evaporate all the solvent before the spectra were collected.

**Transmission electron microscopy (TEM):** 10 µL of ethanol dispersion of purified samples were dropped onto carbon-coated 200-mesh copper grids, followed by evaporation of the solvent. Images were then obtained with a transmission electron microscope (JEOL JEM-1011).

**Photoluminescence (PL) emission measurement:** PL emission of ethanol-dispersed purified samples were all measured in a cuvette with a FluoroMax-4 spectrophotometer using an excitation wavelength of 450 nm. The same experimental settings were also applied for the photostability measurements except that 100 µL of the according ethanol-dispersed samples were diluted with 3mL acetic acid aqueous solution (10%). The PL quantum efficiency of the pure NRs was calculated by comparing the integrated emission of cyclohexane-dispersed NRs to that of ethanol-dispersed rhodamine 6G (quantum efficiency = 94%).[23] Involved in this method, the absorption spectra of the samples were measured by a Cary 5000 UV-Vis-NIR spectrophotometer (Varian). Quantum efficiencies of the ethanol-dispersed NRs/SiO$_2$ were then estimated based on the ratio between its integrated emission to that of the cyclohexane-dispersed NRs.

**Lifetime measurement:** Fluorescence lifetime measurements were performed using a home-built confocal laser microscope. For excitation, the beam of a 448 nm pulsed diode laser with a repetition rate of 20 MHz (Pil044X, EIG2000DX, A.L.S. GmbH) was focused by an objective (10×, Melles Griot, 0.25 NA) on the inner glass surface of a cuvette that contains the solution samples. Light emitted from the solution samples was collected by the same objective and directed through a 532 nm long-pass filter and finally to a single-photon avalanche diode (PDM Series, Micro Photon Devices).

# Results and Discussion

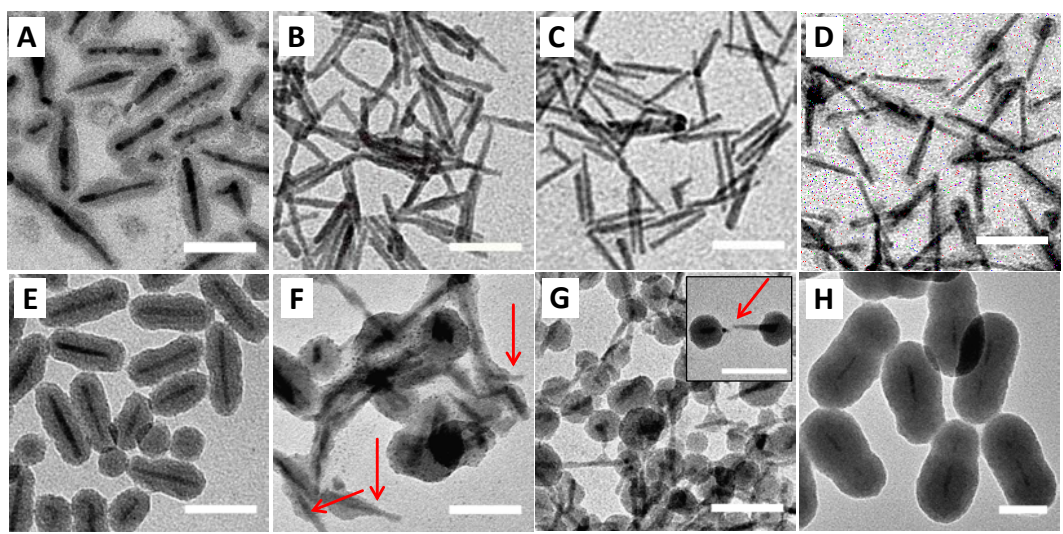

**Figure 1.** TEM images of NRs/SiO$_2$ particles prepared by using (A, E) 60 μL TMOS, (B, F) 82 μL TEOS, (C, G) 100 μL TPOS, and (D, H) 400 μL TEOS as silica precursor after reaction times of (A to D) 1h and (E to H) 24h. The insets of panel G representatively show the breakage of a dumbbell structure. All the scale bars correspond to 50nm.

Figure 1 depicts transmission electron microscopy (TEM) images of the samples synthesized with TMOS (A, E), TEOS (B, F and D, H)), and TPOS (C, G) as described in the experimental section. Here the top and bottom row of images represent structures that have been obtained after 1 h and 24 h reaction time, respectively. The images A-G have been obtained from samples for which an amount of about 0.4 mmol of the respective silane was used in 32.48 mL solution of NRs, Igepal, ammonia, and cyclohexane. The structures of panels D and H were prepared by using a rather large amount of 1.8 mmol TEOS during the synthesis. In the latter case, thick (30 nm) SiO$_2$ shells were formed after 24 h reaction time, as shown in Fig. 1 (H). Moreover from panel D it can be seen that the SiO$_2$ growth process started at the ends of the NRs, which leads to dumbbell-like structures. Both these observations reproduce the results described by Pietra and coworkers.[4]

Since 1.8 mmol (about 400μL) of TEOS results in 30 nm thick silica shells, the use of about 0.4 mmol TEOS should yield a thickness of about 17 nm according to a rough calculation. However, the result is contradictory to such anticipation. As shown in Fig. 1 (B), after 1 h reaction time, no obvious silica at all can be seen from the TEM images. After 24 h growth the silica coverage is very irregular. Moreover the particles are strongly crosslinked with no clear boundaries between individual NRs, as shown in Fig. 1 (F). Besides, as indicated by the red arrows, some structures with silica spheres on a tip occur, suggesting that the silica growth happens preferentially at the tips of the NRs. Summarizing the TEM images in Fig. 1 (B) and (F), we conclude that the $SiO_2$ shells cannot be thinned by simply reducing the concentration of TEOS.

Using TPOS instead of TEOS, but with the same molar amount, yields to structures as shown in Fig. 1 (C, G). Here again, $SiO_2$ formation is hardly observed on the NRs after 1 h (Fig. 1 (C)), however, after 24 h almost perfect $SiO_2$ spheres are formed at the tips of the NRs, with the waist of the NRs remaining totally bare (Fig. 1 (G)). Obviously, the tendency to form such dumbbell-like structures is much more pronounced for TPOS as compared to TEOS. Interestingly, in this sample, we captured some near-broken dumbbell structures, on which the fracture points are well preserved, as indicated by the red arrows in the insets in Fig. 1 (G). This finding strongly supports Anderson et al.'s assumption that NRs with a single-sphere on only one tip are not formed as an intermediate during the reaction, but as a result of the breakage of the primarily formed dumbbell-like structures.

Figure 1 (A) and (E) finally and most importantly show TEM images of structures that have been synthesized using TMOS instead of TEOS or TPOS, but with the same molar amount as the structures shown in Fig. 1 (B, C, F, G). Obviously the use of TMOS leads to a much more homogeneous coverage with a $SiO_2$ shell as compared to TEOS or TPOS. Already after only 1 h reaction time, the NRs are completely covered with a thin $SiO_2$ shell as shown in Fig. 1 (A). The shells in general display a slightly smaller thickness at the waist of the NRs compared to on the tips, implying a remaining preferential $SiO_2$ growth on the tips. However, the preferential growth is by far not as distinct as in the case of TEOS or TPOS. After 24 h reaction time, each of the NRs is surrounded by a uniform $SiO_2$ shell with an average thickness of 8 nm.

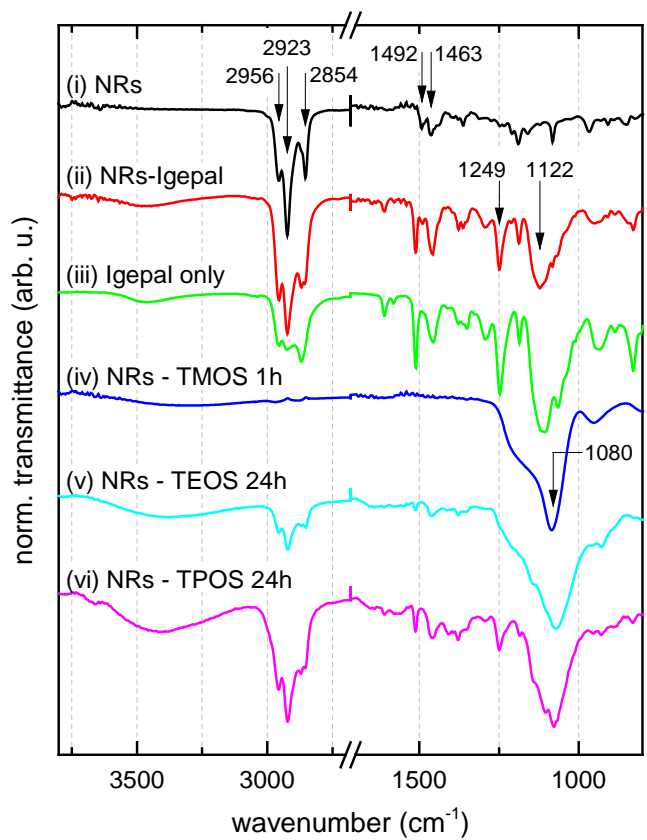

**Figure 2.** FT-IR spectra of, from top to bottom, (i) pure NRs, (ii) Igepal-functionalized NRs, (iii) Igepal only, (iv) 1 h TMOS-treated NRs, (v) 24 h TEOS-treated NRs, and (vi) 24 h TPOS-treated NRs.

Overall, the TEM images in Fig. 1 reveal that the formation of $SiO_2$ shells is faster and more uniform with decreasing alkyl-chain length of the tetra-alkoxysilanes. To gain information about the ligands covering the surface of the particles, we performed FTIR spectrocopy on samples of pure NRs (i) before Igepal functionalization and (ii) after Igepal functionalization but before the silanization with any tetra-alkoxysilane. We further measured (iii) pure Igepal as used in the synthesis as well as silanized NRs obtained using (iv) TMOS, (v) TEOS, and (vi) TPOS by FTIR. Corresponding spectra of the respective samples after washing and removal of excess ligands, and evaporating the solvents are shown in Fig. 2 from top to bottom. For comparison all spectra are offset and normalized to the maximum decrease in transmission. For (i) NRs without further functionalization, the most pronounced features are the strong asymmetric stretching vibration mode of $CH_3$ at 2956 cm$^{-1}$, together with the strong asymmetric and symmetric stretching vibration modes of $CH_2$ at 2923 cm$^{-1}$ and 2854 cm$^{-1}$, respectively. Moreover, two less intense peaks originating from the asymmetric in-plane and symmetric rocking mode of $CH_3$ appear at 1492 cm$^{-1}$ and 1463 cm$^{-1}$, respectively. All the features are assigned to the alkyl chains of the phosphonic acids capping the NRs' surface. In the spectrum of (ii) Igepal-functionalized NRs two additional bands located at 1249 and 1122 cm$^{-1}$ occur, which originate from the asymmetric stretch of –$CH_2$–O–$CH_2$– in Igepal. In the spectrum of (iii) pure Igepal the –$CH_2$–O–$CH_2$– stretching-vibration band at 1122cm$^{-1}$ exhibits a much stronger intensity than the bands from the methyl groups and alkyl chains. In contrast, for the Igepal-functionalized NR sample, the methyl-groups and alkyl-chain bands are slightly more pronounced than the –$CH_2$–O–$CH_2$– band. This indicates that the signal from methyl stretching bands in the Igepal-NRs sample originates from both Igepal and phosphonic acids. Obviously, even after washing and removing of excess ligands, both types of ligands, i.e., Igepal molecules and phosphonic acids are still present on the surface.

For the (iv) TMOS treated NR sample, after only 1 h reaction time, the formation of $SiO_2$ can be identified in the corresponding FTIR spectrum by its characteristic band at 1080 cm$^{-1}$ from the asymmetric vibration of Si–O. The total absence of the methyl stretching bands in the range 2800cm$^{-1}$ to 3000cm$^{-1}$ indicates that neither phosphonic-acid nor Igepal molecules are present in the washed and dried samples. Since the particles are homogeneously covered by $SiO_2$ in this case, it shows that the ligands, which might still be present during the $SiO_2$-shell growth, can be easily washed away from the $SiO_2$ surface. Silica bands are also observed in (v) TEOS- and (vi) TPOS-treated samples, but the bands from the alkyl groups still coexist with the $SiO_2$ bands, even after 24 h reaction time. In combination with the TEM images in Fig. 1 (F) and (G), it is reasonable to assign the alkyl signals to the residue phosphonic acids and Igepal ligands strongly bound to the waist part of the NRs that is not covered by silica.

Based on the FTIR spectra in combination with the TEM investigation, in the following, the formation of silica shells will be discussed with respect to growth models previously described in the literature. The general process of silica growth is, as described in the introduction, already well known. However, the detailed mechanism of silica growth onto hydrophobic nanoparticles exploiting the microemulsion method is still under discussion.[4,12–14,17,24,25] One key question is how individual hydrophobic nanoparticles can be incorporated inside the hydrophilic cores of reversed micelles formed by Igepal molecules in a nonpolar solvent. The exact mechanism seems to strongly depend on the size and the surface composition of the nanoparticles, but in particular also on the chemical nature of the original ligands attached to the nanoparticles. Reports for nanostructures coated with TOPO,[12] with amines,[14,24] with phosphonic acids,[4] and with carboxylic acids[24] propose ligand-exchange processes in which the original ligands are substituted by Igepal and/or partly hydrolyzed TEOS molecules before the nanostructures are incorporated into reversed micelles and the $SiO_2$ growth continues. Pietra et al. proposed for the formation of NR-$SiO_2$ dumbbell-like structures, as reproduced in our work, a mechanism after which the initial $SiO_2$ coverage at the tips of the NRs is due to a favorable ligand exchange with Igepal at the tips because of atomic and hence electronic, or geometric anisotropy.[4]

Based on these models we propose a growth mechanism as sketched in Fig. 3. Starting point is a NR surrounded by phosphonic-acid (mainly ODPA) ligands. The high curvature of the NR surface and the crystallographic facets result in a less compact distribution of original ligands at the tips (cf. Fig. 3 (A)) and thus lead to a smaller steric hindrance of polar Igepal chains to exchange the phosphonic acids. After the addition of Igepal, the original ligands are partially exchanged, as proven by FTIR measurements. Importantly, even though the ligand exchange is expected to be more effective at the tips of the NRs, we assume that Igepal molecules are also located at the sidewalls of the NRs, because NRs covered only with $SiO_2$ spheres at the tips (e.g. after TPOS coverage, see Fig. 1 (G)) show significant IR bands of Igepal ligands, for example an IR band at 1249 cm$^{-1}$ in spectrum (vi) in Fig. 2. This spatial inhomogeneity of the ligand exchange is sketched in Fig. 3 (B). Upon addition of water, ammonia and the tetra-alkoxysilane, the silanization process is triggered. This process is strongly influenced by the size of the alkoxysilane molecule used, in combination with the respective hydrolysis rate. We assume that the smaller TMOS molecules can more efficiently reach the reactive region close to the NR surface with only little steric hindrance (cf. Fig. 3 (C1)), while TEOS (or the even larger TPOS) molecules effectively only reach the regions close to the tips, because of the curvature-reduced steric hindrance (cf. Fig. 3 (C2)). In the reactive regions close to the NR surface, the hydrophilic part of the Igepal molecules provide a water-based environment and the hydrolysis of the tetra-alkoxysilane, if available, can take place as described in eq. (1). Consequently, for TMOS the nucleation of $SiO_2$ on the NR surface and the subsequent $SiO_2$ network formation, as described in eqs. (2-3), can take place all over the NR's surface (cf. Fig. 3 (D1)), while for TEOS (and TPOS) the silica formation occurs only at the tips of the NR (cf. Fig. 3 (D2)). As also verified by the FTIR measurement, in both cases the silica growth leads to a detachment of both Igepal and phosphonic-acid ligands from the semiconductor surface. However, it is believed that the detached Igepal ligands keep surrounding the NRs, with the hydrophilic head facing to the silanized NR surface while the hydrophobic tail points to the cyclohexane "oil" phase. In this condition, the individual micelles are maintained, ensuring the continuity and homogeneity of the whole "water-in-oil" reaction system. In the case of TMOS-prepared structure, regular and uniform Igepal micelles are formed around the NRs (cf. Fig. 3 (E1)). For TEOS-prepared structures, the micelles only form around the silica spheres at both tips of the NRs; between these spheres, original phosphonic acid as well as Igepal ligands are still attached on the NR surface (cf. Fig. 3 (E2)). As already mentioned, these ligands are the reason for the alkene bands visible in the FTIR spectra of TEOS- (or TPOS-) prepared silica-coated NRs as shown in Fig. 2. Note that the rate of silica shell formation is also governed by the hydrolysis rate, which is faster for TMOS compared to TEOS (or even TPOS).[21] As a consequence, the growth of a complete shell using TMOS, like sketched in the left column of Fig. 3, is much faster than the formation of the corresponding dumbbell-like structures sketched in the right column for the use of TEOS.

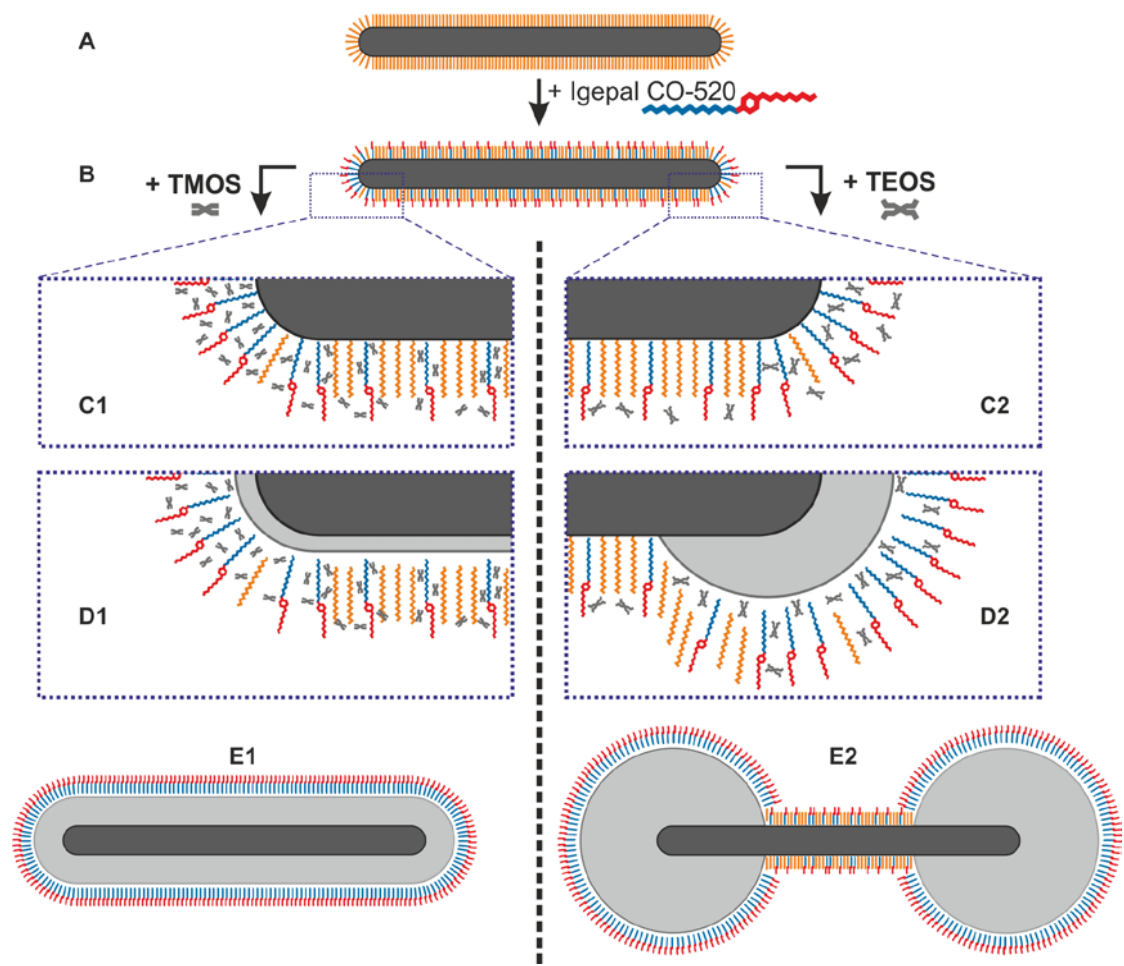

**Figure 3.** Sketch of the silica growth process. (A) NR with original phosphonic-acid ligands attached. (B) Inhomogeneously exchanged ligands after addition of Igepal. (C) Effect of the steric hindrance for (C1) TMOS and (C2) TEOS. (D) Growth of (D1) a thin homogeneous silica shell for the use of TMOS and (D2) silica spheres at the NR tips for the use of TEOS. (E1) Silica coated NRs in an Igepal micelle. (E2) NR-SiO$_2$ dumbbell-like structure with micellar Igepal at the tips and partly exchanged ligands still attached at the NR center.

The growth mechanism sketched in the right column of Fig. 3, i.e. for the use TEOS, which leads to dumbbell-like structures, is similar to the one proposed by Pietra et al. in Ref. [4]. Interestingly, the same group recently also reported on the formation of ultrathin silica shells around anisotropic semiconductor nanocrystals even with the use of TEOS.[13] Here, an extremely low concentration (0.6%) of ammonium-hydroxide solution led to a strongly decreased hydrolysis rate of the TEOS (see eq. (1)). It was then argued that the only partial hydrolysis of TEOS leads to a more drastic decrease of the silica growth rate compared to the nucleation rate. As a consequence, the authors expected that the original capping ligands of the NRs were exchanged by (partly) hydrolysed TEOS monomers before stable silica networks could develop. Interestingly, in our case, with the use of TMOS instead of TEOS, we somehow go the opposite way: the hydrolysis rate is strongly enhanced; we expect all TMOS molecules to be fully hydrolyzed very quickly. From this, one would expect that both the nucleation and the silica growth rate are increased. In addition to the above discussed size effect, these changes in rates can support the TMOS-induced homogeneous silica growth when the ratio of nucleation and growth rates is increased for the use of TMOS compared to TEOS or even TPOS. In fact, the influence of the length of the alkoxy chain of the tetra-alkoxysilane on the formation of (empty) silica spheres has been extensively investigated.[26–29] The studies demonstrate that the use of TMOS can create a larger number of silica spheres with smaller size and narrower size distribution compared to TEOS, and TPOS, indicating a higher nucleation-to-growth-rate ratio for TMOS compared to the other tetra-alkoxysilanes. Even though these studies deal with the homogeneous nucleation of silica, the general trend might still be valid also for our work, in which the heterogeneous nucleation of hydrolysate monomers on the surface of NRs takes place.

In the following we want to analyze the fluorescence properties of the samples discussed in terms of their FTIR spectra in Fig. 3, which are of major importance for their further use for all kinds on applications. Figure 4 shows (A) the PL emission spectra and (B) the corresponding decay curves, all measured on the purified samples dispersed in ethanol. First of all we stress that the samples dispersed in ethanol (black curve in Fig. 4 (A)) display a lower PL emission compared to their counterparts in the reaction mixture in which cyclohexane is the solvent (not shown here). The quantum yield of the pure NRs in cyclohexane was measured to be about 35%. It decreased after their transfer to ethanol by a factor of 10, i.e., to a value of 3-5%, due to a solvent effect.[30] While the addition of the Igepal solution leads to a further decrease of the PL intensity, the following surface modification with silica again increases the PL intensity in the sequence TPOS, TEOS, and TMOS. The latter sample displays a PL emission more than twice as intense as that of the pure NRs, i.e., a quantum yield of 8-15%. Importantly, the sequence of PL intensities for the different samples is closely corresponding to changes in the PL lifetimes. In general, the lifetime is drastically reduced for the NR-Igepal sample as well as for the TEOS- and TPOS-treated samples, compared to the pure NR sample. Most noteworthy, the TMOS-treated sample exhibits a lifetime even longer than the pure-NR sample. All decay curves are more complex than mono-exponential. The decay curves of the pure NRs and of the TMOS-prepared NR-SiO$_2$ sample reveal long-lived decay channels with lifetimes larger than the 50 ns time interval between two excitation-laser pulses, as can be seen in the above-baseline signal before time zero in Fig. 4 B. The decay curves of the other samples essentially reach the baseline within 50 ns. An in-depth analysis of the fluorescence decay is beyond the scope of this paper, but in order to give a rough measure of the different decays of each sample, the 1/e lifetime may be used even though all the decays are not mono-exponential. These lifetimes are slightly below 6 ns for the pure NR and the TMOS-prepared NR-SiO$_2$ sample, while the other samples exhibit lifetimes of about 1.3 ns. The fluorescence decay measurements reveal that the partial exchange of the original phosphonic acids by Igepal leads to a weaker protection of the NR with respect to the ethanol solvent, thus new non-radiative recombination channels open up and lead to a shortening of the measured lifetime. A partial coverage of the NRs by silica, as is the case for the TEOS- and TPOS-prepared structures (cf. Fig.1 (F) and (G)), does not significantly close these non-radiative channels. Strikingly, the thin silica shell prepared with TMOS obviously strongly protects the NRs against the ethanol solvent again; the corresponding lifetime is even slightly longer than the one of the pure phosphonic-acid-capped NRs in ethanol.

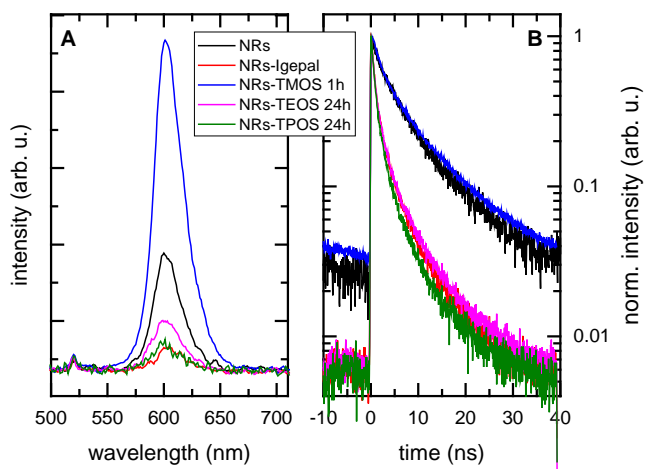

**Figure 4.** (A) PL emission spectra and (B) corresponding decay curves of pure NRs (black), Igepal-functionalized NRs (red), 1 h TMOS-treated NRs (blue), 24 h TEOS-treated NRs (magenta) and 24 h TPOS-treated NRs (green). All samples have been excited at a wavelength of 449 nm. The samples have been washed once and redispersed in ethanol.

Another major aspect of the silica coverage is the effectivness of chemical passivation to ensure long time fluorescence stability. In order to evaluate the shielding ability of the thin silica shell prepared by using TMOS, we compare the temporal stability of the PL emission of the silica-coated structures upon exposure to acidic solution. Previous research has confirmed that acidic solution efficiently quenches the fluorescence of bare NPs.[5] Figure 5 shows the time evolution of the integrated PL peaks of NRs with a 4 nm SiO$_2$ shell prepared with TMOS (red circles) to that of NRs with a 30 nm SiO$_2$ shell prepared with TEOS (black squares), where both samples were dispersed in acetic acid aqueous solution (10%, pH=2) over 5 h. The data points are normalized to the respective initial value at time zero. Note that the TMOS-prepared sample corresponds to the sample as shown in Fig. 2 (A) while the TEOS-prepared sample is shown in Fig. 2 (H). Both samples first exhibit a decrease in emission intensity; after 20 min, both samples emit only 70% to 80% of the initial PL intensity. After that, however, the TMOS-prepared structures show a nearly constant intensity level between 70% and 80% of the initial value, while the TEOS-prepared structures exhibit a drastically decreasing intensity. After 2 hours, the emission intensity of these structures has dropped already below 30%, further decreasing below 20% during the following 3 hours.

We assume that the initial fast decrease in PL emission intensity for the TMOS sample is caused by a subensemble of NRs with an incomplete silica shell, whose emission is quickly quenched by the acidic acid. The remaining NRs exhibit a superior shielding indicating a complete silica shell, even though the shell is only 4 nm thin. The strong decrease in PL emission intensity for the TEOS-prepared structures cannot be explained with a larger subensemble of NRs with an incomplete shell since coating with 30 nm of silica, compared to 4 nm as for the TMOS structures, should decrease the number of NRs with an incomplete shell. In fact, the strong and the continuous decrease of PL intensity over a long time range indicate that the thick TEOS-prepared shells are more porous than the thin TMOS-prepared counterparts. This difference in porosity can be easily explained with the aforementioned difference in hydrolysis rate between TMOS and TEOS. As mentioned above, we expect all TMOS molecules to be fully hydrolyzed very quickly. This means that the TMOS-initiated $\equiv$Si−O−Si$\equiv$ matrix is formed via the interconnection between fully hydrolyzed silica species and is thus "pure", as sketched in the upper scheme in Fig. 5. On the other hand, TEOS as the precursor only partially hydrolyzes and the TEOS-initiated $\equiv$Si−O−Si$\equiv$ matrix may contain a larger amount of unreacted alkyl chains embedded inside during the build-up (see lower scheme in Fig. 5). With the reaction going on, the embedded alkyl chains might get gradually hydrolyzed and released in form of ethanol, resulting in a higher porosity of the finally formed silica shell, and thus in a weaker passivation of the NRs with respect to acetic acid.

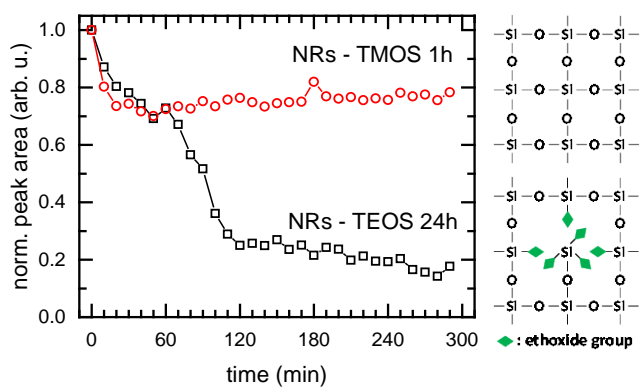

**Figure. 5.** Evolution of the integrated area of PL emission peak of the NRs/SiO$_2$ prepared by using 60 μL TMOS after 1 h reaction time (red circles, cf. Fig. 1 (A)) and 400 μL TEOS after 24 h reaction (black squares, cf. Fig. 1 (H)) dispersed in acetic acid aqueous solution (10%) over 5 h. The schemes illustrate the TMOS (top) and TEOS (bottom) initiated $\equiv$Si−O−Si$\equiv$ matrix.

## Summary and Conclusions

In summary, we systematically investigated the effect of the alkyl-chain length of tetra-alkoxysilanes on the silica formation on CdSe/CdS NRs applying the microemulsion technique with Igepal as the nonionic tenside. We find an enhanced spatial homogeneity of the silica growth with decreasing alkyl-chain length, from TPOS over TEOS to TMOS. Strikingly, the use of TMOS leads to a homogeneous and continuous encapsulation of NRs into a thin silica shell within only one hour reaction time. FTIR studies reveal that, during the synthesis, the Igepal molecules partly exchange the original phosphonic-acid ligands on the NRs. We suggest that it is the lower steric hindrance of TMOS molecules compared to TEOS and TPOS that allows the hydrolysis of the tetra-alkoxysilane to take place not only close to the tips but also along the sidewalls of the NRs. Consequently, here, the silica nucleation can take place homogenously on the NR surface. The faster hydrolysis rate for TMOS compared to TEOS or TPOS directly leads to a more rapid growth of the silica shell. A faster hydrolysis rate, as for TMOS, also promotes the full hydrolysis of all alkoxy groups of the silane molecule and the polymerization of fully hydrolyzed siliane molecules should lead to less porous silica networks as compared to only partly hydrolyzed molecules. In fact, the TMOS-prepared thin ($\leq 5$ nm) silica shell shows an even superior shielding ability against acidic environment compared to the TEOS-prepared 30 nm thick one. The approach of using TMOS presented here may also open a window for rapidly creating thin and dense silica shells on other types of anisotropic nanostructures with hydrophobic surfaces.

## Acknowledgements

T.K. acknowledges funding from the European Union's Horizon 2020 research and innovation programme under the Marie Skłodowska-Curie grant agreement No. 656598. We acknowledge financial support from the German Research Foundation (DFG) via the Cluster of Excellence "Centre for Ultrafast Imaging" (CUI).

Table of Contents/Abstract Graphic

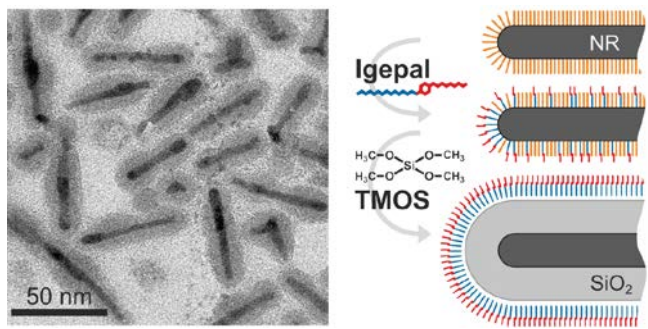